\documentclass[aip, rsi, amsmath, amssymb, reprint]{revtex4-1}

\usepackage{graphicx}
\usepackage{dcolumn}
\usepackage{bm}
\usepackage{units}

\begin{document}

\title{Coherent control using kinetic energy and the geometric phase of a conical intersection}

\author{Chelsea Liekhus-Schmaltz}
\email{cliekhus@stanford.edu}
\affiliation{ Stanford PULSE Institute, SLAC National Accelerator Laboratory\\
2575 Sand Hill Road, Menlo Park, CA 94025}
\affiliation{Department of Physics, Stanford University, Stanford, CA 94305}

\author{Gregory A. McCracken}
\affiliation{ Stanford PULSE Institute, SLAC National Accelerator Laboratory\\
2575 Sand Hill Road, Menlo Park, CA 94025}
\affiliation{Department of Applied Physics, Stanford University, Stanford, CA 94305}

\author{Andreas Kaldun}
\affiliation{ Stanford PULSE Institute, SLAC National Accelerator Laboratory\\
2575 Sand Hill Road, Menlo Park, CA 94025}
\affiliation{Department of Physics, Stanford University, Stanford, CA 94305}

\author{James P. Cryan}
\affiliation{ Stanford PULSE Institute, SLAC National Accelerator Laboratory\\
2575 Sand Hill Road, Menlo Park, CA 94025}

\author{Philip H. Bucksbaum}
\affiliation{ Stanford PULSE Institute, SLAC National Accelerator Laboratory\\
2575 Sand Hill Road, Menlo Park, CA 94025}
\affiliation{Department of Physics, Stanford University, Stanford, CA 94305}
\affiliation{Department of Applied Physics, Stanford University, Stanford, CA 94305}

\date{\today}

\begin{abstract}
Conical intersections (CI) between molecular potential energy surfaces with non-vanishing non-adiabatic couplings generally occur in any molecule consisting of at least three atoms. 
They play a fundamental role in describing the molecular dynamics beyond the Born-Oppenheimer approximation and have been used to understand a large variety of effects, from photofragmentation and isomerization to more exotic applications such as exciton fission in semiconductors. 
However, few studies have used the features of a CI as a tool for coherent control.  
Here we demonstrate two modes of control around a conical intersection.  
The first uses a continuous light field to control the population on the two intersecting electronic states in the vicinity of a CI.  
The second uses a pulsed light field to control wavepackets that are subjected to the geometric phase shift in transit around a CI.
This second technique is likely to be useful for studying the role of nuclear dynamics in electronic coherence phenomena.  
\end{abstract}

\maketitle

\section{Introduction}

Quantum calculations of transient-state photochemistry and photobiology rely on approximations to deal with the large number of interactions between nuclei and electrons~\cite{klessinger_excited_1995, allen_computational_2009}.
The Born-Oppenheimer framework often serves as a first attempt to simplify these problems by invoking the separation of time scales between nuclear and electronic motion~\cite{born_zur_1927}. 
This framework consists of two main parts:  
First, it describes the total molecular wavefunction as a product of the nuclear wavefunctions and the adiabatic electronic eigenstates, which are calculated at fixed nuclear geometries.  
Second, this framework invokes the adiabatic approximation which states that nuclear motion does not couple different electronic states, because the derivatives of the electronic wavefunction with respect to the nuclear coordinates are small.

The adiabatic approximation breaks down if two or more eigenenergies approach degeneracy.  
In this case, as shown by Von Neumann and Wigner~\cite{von_neumann_uber_1929}, any molecule with $N=3$ or more atoms and $3N-6$ degrees of freedom will have a $3N-8$ dimensional seam of degeneracy.  
The remaining two degrees of freedom define a 2-D subspace called the ``$g-h$" plane in which the adiabatic potentials look like two cones joined at a point -- hence the name ``conical intersection'' (CI)~\cite{baer_beyond_2006,worth_beyond_2004,yarkony_diabolical_1996,domcke_conical_2004}.  
Near the CI the electronic states change rapidly with small changes in the configuration of the nuclei.
This leads to nonadiabatic coupling terms that increase as the inverse of the energy splitting between the two surfaces and diverge exactly at the degeneracy.
These infinities can be avoided by moving to what has been called the diabatic picture.
In the context of CIs, the diabatic picture is defined as a rotation of the adiabatic states at each point in nuclear configuration space~($R$), such that the expectation value of the nuclear kinetic energy operator on the electronic states in the rotated `diabatic' basis is zero.
As a result of the rotation, the couplings between electronic states enter the Hamiltonian as a local off diagonal terms in the diabatic potential $W$,  rather than the derivative coupling terms in the adiabatic picture. 
In practice, however, it is not possible to write down such a rotation, but one can approximate the diabatic picture by making a Taylor expansion of the adiabatic potentials and electronic eigenstates around the CI~\cite{van_voorhis_diabatic_2010}.
The truncated expansion leads to the first-order diabatic potential:
\begin{equation}
	W =
    \begin{bmatrix}
    	W_{11}(R) & \lambda y\\
        \lambda y & W_{22}(R),
    \end{bmatrix} \label{Potential}
\end{equation}
where the diagonal terms, $W_{ii}(R)$ are Taylor expansions of the adiabatic eigenvalues evaluated about the CI~($R=R_0$).
The off diagonal terms are related to the derivative couplings between the adiabatic states with respect to the displacement, $y$, along the $h$-direction. 
The adiabatic potential energy surfaces in the vicinity of the CI are the eigenvalues of this diabatic potential~($W$). 
Rotation of the nuclear kinetic energy operator back into the adiabatic picture generates the diverging nonadiabatic coupling terms.  

Experimental and theoretical studies have shown that CIs provide a pathway for ultrafast non-radiative population transfer between electronic states~\cite{teller_crossing_1937, levine_isomerization_2007,domcke_conical_2004,domcke_conical_2011}.
In addition, recent papers have considered how non-resonant light-fields can control the dynamics around a CI~\cite{kowalewski_catching_2015, kim_control_2012,sussman_dynamic_2006,townsend_stark_2011, sussman_nonperturbative_2005}.  
In this work we will consider how a weak, resonant, linearly polarized laser field can be used to control the dynamics near a conical intersection. 

We consider the effect of a light field on a triatomic molecule with linear vibronic coupling and dipole-coupled excited states.
We have identified two types of control, each named for the property that allows us to control the wavepacket.
``Kinetic energy control'' is exerted by applying a continuous light field for the duration of the molecular dynamics.  
The wavelength of the light allows us to control the population distribution on each of the electronic states by changing the kinetic energy of the wavepacket near the CI.
This control mechanism is depicted schematically in the top portion of Fig.~\ref{fig:KEC}.
The other control mechanism depicted in the bottom portion of Fig.~\ref{fig:GPC}, we have termed `'`Geometric phase control''.
This technique uses a light pulse to couple the two coherent wavepackets on the upper and lower diabatic states that are produced in traversing the CI.
The geometric phase accumulated in transit around the CI influences the interference and leads to an asymmetric wavepacket~\cite{althorpe_effect_2008}.  
The asymmetry can be changed depending on the phase difference between the light pulse and the wavepacket coherence.  

\section{Three Step Simulation}

\begin{figure}
  	\includegraphics[width =\columnwidth]{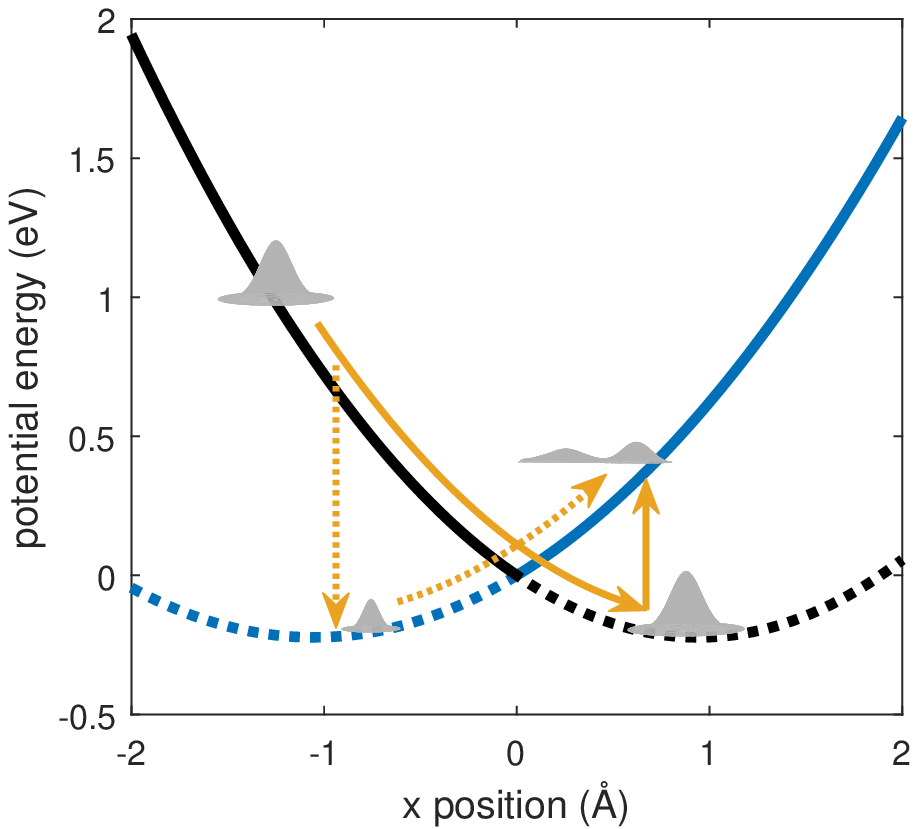}\\
   	\includegraphics[width=\columnwidth]{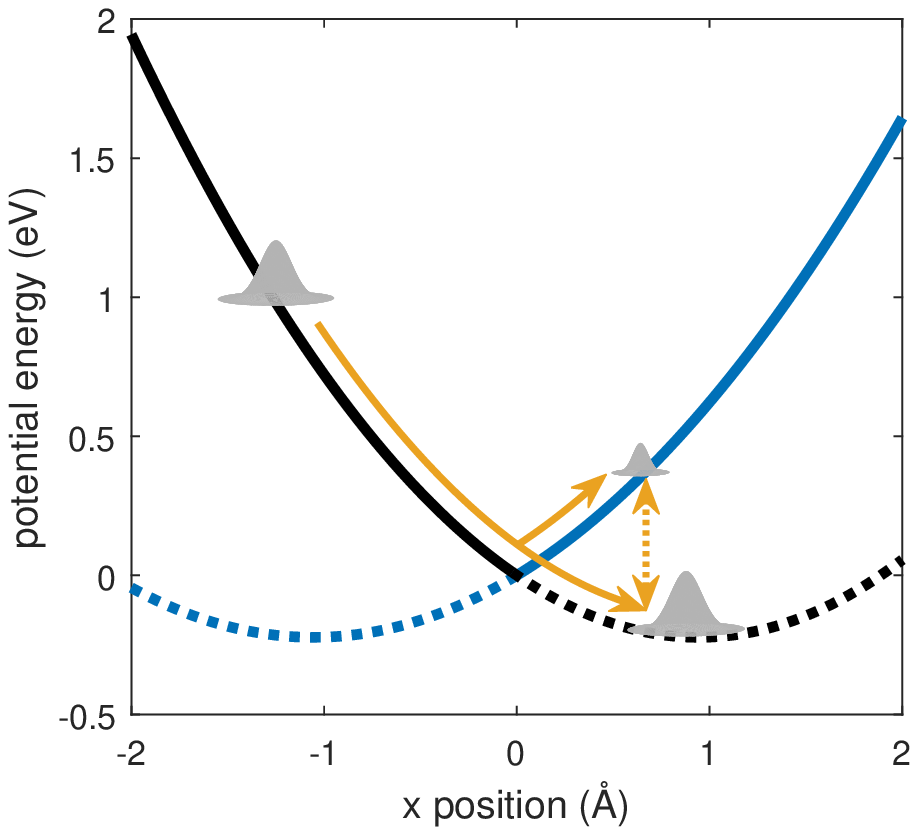}
    \caption{Diabatic potential energy surfaces~(black~and~blue), which are the diagonal elements of eqn.~\ref{Potential}, and adiabatic potential energy surfaces~(dashed~and~solid) for $y=0$. The upper panel is an illustration of ``kinetic energy control.'' Here a continuous light wave creates a pathway indicated by the yellow dashed lines. The wavepacket~(gray) gains kinetic energy on the upper-state then the laser-field allows the wavepacket to transition from the upper-state to the lower-state. Depending on the time of the transition (or the wavelength of the control field) the wavepacket velocity when entering the CI will change, altering the non-adiabatic transition probability. The lower panel shows an illustration of the ``geometric phase control'' concept. The CI acts as a quantum beam-splitter and creates coherent wavepackets on both the upper and lower states. The laser field couples these two wavepackets and the shape of the wavefunction depends critically the relative phase of the laser pulse.}
    \label{fig:KEC}
    \label{fig:GPC}
    \label{fig:potentials}
\end{figure}

The diabatic potential we consider is given by $W$ in eqn.~\ref{Potential} with diagonal elements:
\begin{equation}
	W_{ii}(x,y,z) = \kappa_{x,i} (x-x_i)^2 + \kappa_{y,i} y^2  +\kappa_{z,i} z^2 - E .
\end{equation}
The diagonal terms are shown as black and blue lines in Figure~\ref{fig:potentials}.  
The corresponding adiabatic potential energy surfaces are generated by diagonalizing the diabatic potential matrix and are also shown in Fig.~\ref{fig:potentials}.

The wavepacket propagates on a $\unit[3]{\text{\AA}}$~cube containing $128^3$ grid points with time steps of $0.1$ femtoseconds.  
The initial wavepacket is a ``coherent state" in the parabolic potential~(i.e. a non-spreading Gaussian wavepacket)~\cite{tannor_introduction_2007} with an initial position of $x_0=-1.32x_2$.
We adjust the initial position of the wavepacket such that passage through the CI is nearly total, i.e. in the absence of a coupling laser all the population is transferred to the adiabatic ground state.  
The rest of the constants used in the simulation are shown in Table \ref{tab:constants}.  
These parameters correspond to a wavepacket that moves from its initial position and through the CI in about $\unit[15]{fs}$.  
The wavepacket will reach the outer-turning point and return to the CI after $\sim\unit[30]{fs}$.

\begin{table}[t!]
\centering
\begin{tabular}{|c|c|}
    	\hline
    	Parameter & Value\\\hline
        $x_2$ & $\unit[0.944]{\text{\AA}}$\\
        $x_1$ & $-1.118x_2$\\
        $x_0$ & $-1.32x_2$\\
        $\kappa_{x,2}$ & $\unitfrac[0.25]{eV}{\text{\AA}^2}$\\
        $\kappa_{x,1}$ & $0.8\kappa_{x,2}$\\
        $\kappa_{y,1}=\kappa_{y,2}=\kappa_{z,1}=\kappa_{z,2}$ & $0.4\kappa_{x,2}$\\
        $\lambda$ & $\unitfrac[0.0424]{\kappa_{x,2}}{x_2}$\\
        $E$ \tiny{(this could also be set to 0)}& $\kappa_{x,1}\cdot x_1^2$\\
        mass & $\unit[1]{amu}$\\\hline
\end{tabular}
\caption{Model parameters}
\label{tab:constants}
\end{table}
    
Normally, the simple split step method is sufficient to simulate the dynamics under the diabatic potential, $W$ \cite{kosloff_fourier_1983}; however in order to include the dipole coupling we need to add in another step.  
The linear coupling Hamiltonian is defined in the diabatic picture and the dipole coupling matrix is defined in the adiabatic picture.  
Therefore, it is necessary to either rotate the dipole coupling matrix into the diabatic picture, or to rotate the nuclear wavefunction into the adiabatic picture on each time step in order to apply the dipole coupling.  
We propagate the initial state in the diabatic picture according to:  
\begin{eqnarray}
\psi(t+\Delta t)&=&U(t,\Delta t)\psi(t) \\
U(t,\Delta t)&=& U_A e^{-i\hat{D}(t)\Delta t} U_A^{\dagger} e^{-i\hat{W}\Delta t} \nonumber \\
& & \times U_k(t) e^{-i\hat{T}_N\Delta t} U_k^{\dagger}(t). 
\label{eqn diabatic propagator}
\end{eqnarray}
$U_A$ is the transformation between the diabatic and adiabatic picture, $U_k(t)$ is the Fourier transform operator from momentum space to configuration space, and $U(t,\Delta t)$ is the time propagator.

Equation \ref{eqn diabatic propagator} has a simple interpretation:  
Begin with a nuclear wavefunction in the diabatic picture,~$\psi(t)$. 
Apply a Fourier transform ~$U_k^{\dagger}(t)$ and the diagonal kinetic energy propagator, $e^{-i\hat{T}_N\Delta t}$, where $\hat{T}_N$ is the nuclear kinetic energy operator and $\Delta t$ is the time step. 
Then apply the inverse Fourier transform,~$U_k(t)$.  
Next, apply the diabatic propagator, $e^{-i\hat{W}\Delta t}$, where $W$ is given by eqn.~\ref{Potential} in the diabatic basis.  
Finally, rotate into the adiabatic picture $U_A^{\dagger}$ and apply the dipole coupling propagator $e^{-i\hat{D}(t)\Delta t}$, where $\hat{D(t)}$ is the time-dependent dipole operator matrix, which is off-diagonal in the adiabatic picture. 
In order to rotate between diabatic and adiabatic pictures the grid should not include the point of the conical intersection, since the rotation matrix is not well defined at this point.

These three operators do not commute with one another, and breaking up the Hamiltonian in this way is an approximation.  
However, as in the original split step method, if the time steps are small enough, the error, $\sigma_e$, is small:
\begin{equation}
	\begin{split}
    	\sigma_{e} =& \frac{1}{2\hbar}\left([\hat{D}(t),\hat{W}]+[\hat{D}(t), \hat{T}_N]+[\hat{W},\hat{T}_N]\right)(\Delta t)^2 \\
        &+ \mathcal{O}((\Delta t)^3).
    \end{split}
\end{equation}

\subsection{Kinetic Energy Control}

\begin{figure}
	\centering
    \includegraphics[width = .5\textwidth]{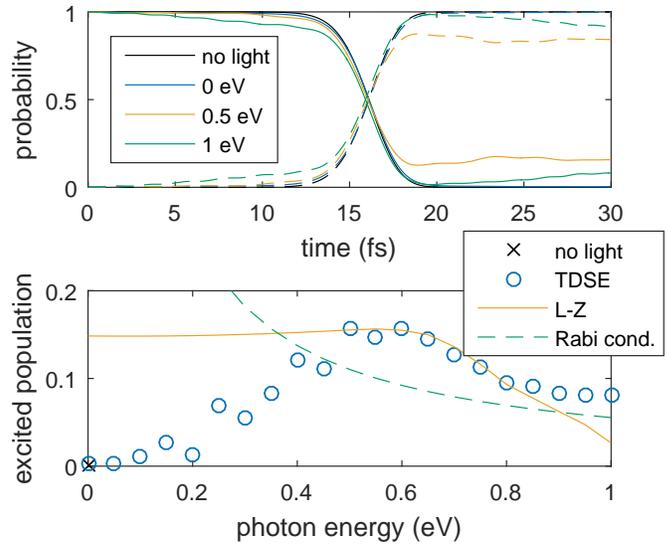}
    \caption{Time-dependent population of the ground adiabatic~(dashed) and excited adiabatic~(solid) states for field strength $F=\unitfrac[10^9]{V}{m}$ (top panel).
	Initially only the excited state is populated, and the linear coupling parameter, $\lambda$, is adjusted such that in the field-free case~(black-curve), nearly all of the population is transferred to the ground-state after passing through the conical intersection.
    The total population of the excited state after $\unit[30]{fs}$ of propagation is plotted as blue circles in the bottom panel~(TDES).
    The oscillation structure below $\sim\unit[0.5]{eV}$ depends on the total propagation time, due to the time-dependent oscillations shown in the top panel.
    These oscillations appear because at such low photon energies, the photoabsorption and nonadiabatic dynamics near the conical intersection cannot be clearly separated, see text. 
    The yellow curve~(L-Z) shows the result of the sequential model described by Eqn.~\ref{TotalProb}, which reproduces the simulation results until the assumption of a sequential process breaks down for low photon energies.
    The dashed curve shows $\nicefrac{\bar{\Omega}}{\left. \Omega(\omega)\right|_{CI}}$, which describes the validity of the sequential model.}
    \label{fig:populations}
\end{figure}

First we consider the effect of a continuous light field on the wavepacket propagation, as illustrated in Figure \ref{fig:KEC} (top panel).  
The simulation is performed with a constant light field amplitude $F = \unitfrac[10^9]{V}{m}$.  
The time-dependent population of both the ground and excited adiabatic states is shown in Figure~\ref{fig:populations} (top panel).  
Figure~\ref{fig:populations}~bottom panel shows the population of these states $\unit[30]{fs}$ after the start of the wavepacket propagation, when the wavepacket starts to turn around on the upper adiabatic state.

The addition of an oscillating light field has a marked effect on the intermediate and final population.  
This is emphasized by the fact that a static electric field (labeled as $\unit[0]{eV}$ in the top panel of Fig.~\ref{fig:populations}) has a negligible effect compared to an oscillating field with a $\unit[0.5]{eV}$ photon energy.  
Figure~\ref{fig:populations}~bottom panel shows that there is an optimal photon energy for maintaining population in the excited state.
Below this optimal photon energy, the excited state yield has an oscillatory structure.
The details of this structure depend on the total wavepacket propagation time, i.e.~the structure would look different if the wavepacket is propagated for $\unit[25]{fs}$ or $\unit[35]{fs}$.
For the field parameters used to create Figure~\ref{fig:populations} ($F=\unitfrac[10^9]{V}{m}$), the optimal photon energy for population transfer is $\sim\unit[0.5]{eV}$, which is approximately half of the energy separation between the ground and excited states at the initial wavepacket position $x_0$.  
We can explain the existence of the optimal photon energy using a semi-classical model. 
Using the diabatic picture, the wavepacket initially propagates toward the conical intersection.
When the wavepacket is in the vicinity of the one-photon resonance between the two states, a portion of the population is transferred to the ground state. The wavepacket then propagates towards the CI on the lower surface as shown by the dashed yellow line in Fig.~\ref{fig:KEC}.
According to the Landau-Zener formula \cite{landau_quantum_1958}, the non-adiabatic population transfer probability, $P_{CI}$, depends on the velocity of the wavepacket in the vicinity of the CI, $v_{CI}$:
\begin{equation}
P_{CI}=\mbox{Exp}\left[-\frac{2\pi V_{UL}^2}{v_{CI}(\omega)}\left[\frac{\partial\Delta E}{\partial x}\right]^{-1}_{CI}\right], \label{LandauZener}
\end{equation}
where $V_{UL}$ is the coupling between the two diabatic states, and $\nicefrac{\partial\Delta E}{\partial x}$ describes the difference in slope of the two states at the CI.
Changing the wavelength of the control field affects where on the excited state surface the transition is made, and this location will affect the velocity of the wavepacket in the vicinity of the CI. 
From equation~\ref{LandauZener}, we see that faster wavepackets are more likely to make non-adiabatic transitions around the conical intersection and thus remain on the diabatic surface as they move through the vicinity of the conical intersection.

In addition to the Landau-Zener formula for the non-adiabatic transition probability, the probability for transitioning to the lower excited states needs to also be considered.
This transition probability will depend on the speed of the wavepacket in the vicinity of the one-photon resonance~($\Delta E\sim\omega$).
The relevant parameters for defining the population transfer by the light-field is related to the non-adiabatic correction for adiabatic passage via a frequency sweep~\cite{tannor_introduction_2007}:
\begin{equation}
P_{L}=1-\mbox{Exp}\left[-\frac{\pi \bar{\Omega}^2 \hbar}{2 v_{x<0}(\Delta E\sim\omega)}\left[\frac{\partial\Delta E}{\partial x}\right]^{-1}_{\Delta E\sim\omega}\right], \label{TransitionProbL}
\end{equation}
where $\bar{\Omega}=\nicefrac{\mu F}{\hbar}$ is the Rabi frequency for electric field strength $F$ and transition dipole moment $\mu$. 
The speed of the wavepacket at resonance, $v_{x<0}(\Delta E\sim\omega)$, is  calculated using a classical trajectory. $\nicefrac{\partial\Delta E}{\partial x}$ describes the rate of change of the energy separation of the two states ($\Delta E$) with in the direction of propagation~($x$).
So far we have described the process depicted in Fig.~\ref{fig:KEC} as the yellow dashed line.
We  must also consider another pathway which leads to population on the excited state after the wavepacket traverses the CI. This pathway is illustrated by the solid yellow line in Fig.~\ref{fig:KEC}. Population that is not transferred at the first point of resonance can be transferred on the other side of the CI since there is a resonance there as well. As defined by the simulation parameters, $P_{CI}=1$ along this pathway. Other pathways that might be considered are higher order and have negligible contributions at these field strengths. 
The total probability, $P_E$, for being in the excited state after $\unit[30]{fs}$ of propagation is given by the sum of the probability for the two different pathways, 
\begin{eqnarray}
P_E &=& \overbrace{P_L*P_{CI}}^{\mbox{yellow dash}} + \overbrace{(1-P_L)P_R}^{\mbox{solid yellow}} 
\label{TotalProb}
\end{eqnarray}
where $P_L$ is defined in eqn.~\ref{TransitionProbL}, $P_{CI}$ is given by eqn.~\ref{LandauZener}, and $P_R$ is the probability for the wavepacket to be photoexcited on the right side~($x>0$) of the conical intersection: 
\begin{equation}
P_{R}=1-\mbox{Exp}\left[-\frac{\pi \bar{\Omega}^2 \hbar}{2 v_{x>0}(\Delta E\sim\omega)}\left[\frac{\partial\Delta E}{\partial x}\right]^{-1}_{\Delta E\sim\omega}\right], \label{TransitionProbR}
\end{equation}
now $v_{x>0}(\Delta E\sim\omega)$ is the speed of the wavepacket on the lower surface of the right side~($x>0$) of the conical intersection, again calculated using a classical trajectory. 

\begin{figure} [t]
    \includegraphics[width = .5\textwidth]{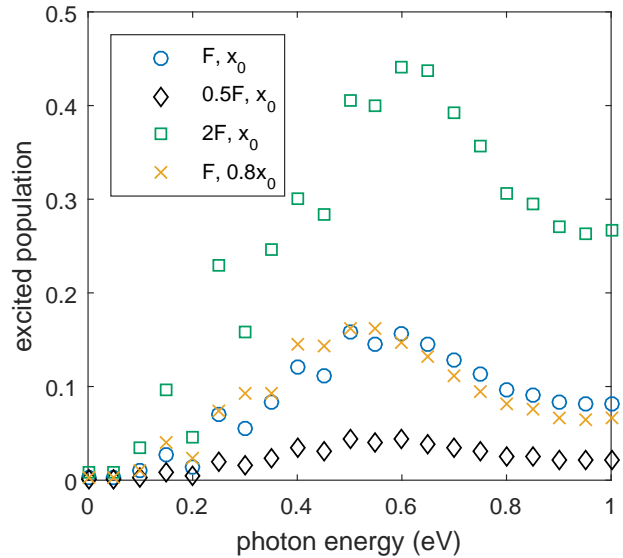}
    \caption{Total population of excited states after $\unit[30]{fs}$ as a function of the control field wavelength, for different simulation parameters: electric field strength, $F$, and initial starting points, $x_0$, for wavepacket. 
    The o's are the same data set as depicted in Fig.~\ref{fig:populations}, featuring a maximum in the population transfer for a photon energy of $\unit[0.5]{eV}$.
    Also shown is the simulation for $F=\unitfrac[0.5\times10^9]{V}{m}$, $x_0=-1.32x_2$~(diamonds), $F=\unitfrac[2\times10^9]{V}{m}$, $x_0=1.32x_2$~(squares), and $F=\unitfrac[10^9]{V}{m}$,$x_0=1.06x_2$~(x's)
    The larger $F$, the more population is transferred.  
    For lower $x_0$ and the corresponding lower wavepacket velocities, the maximal population transfer is shifted to slightly lower photon energies.}
    \label{fig:CompareEV}
\end{figure}

Equation \ref{TotalProb} is averaged over classical trajectories weighted by the initial nuclear wavepacket. 
The resulting probability is plotted in Fig.~\ref{fig:populations} and shows excellent agreement with the simulation for photon energies between $\sim\unit[0.5]{eV}$ and $\unit[0.8]{eV}$. 

The model breaks down at photon energies greater than $\unit[0.8]{eV}$ because eqn.~\ref{TransitionProbL} is only valid for nuclear wavepackets that pass through the resonance, rather than starting at rest at the resonance. 
$P_L$ then overestimates the amount of population transferred due to the slow wavepacket speeds near $t=0$. 
For the low photon energy range, resonant population transfer and transfer through the CI is no longer sequential, as assumed in eqn.~\ref{TotalProb}.  
To quantify the breakdown of this model, we consider the ratio of the Rabi frequency, $\bar{\Omega}$, to the generalized Rabi frequency, $\Omega=\sqrt{\bar{\Omega}^2+(\Delta E-\omega)}$, evaluated at the CI~$(\Delta E=0)$ for each light frequency, $\omega$.
If this ratio is much less than 1~$\left(\nicefrac{\bar{\Omega}} {\left.\Omega(\omega)\right|_{CI}} \ll 1 \right)$, then the control field interaction and the dynamics near the CI can be thought of separately, and the process can be treated sequentially.
We plot the ratio $\nicefrac{\bar{\Omega}}{\left.\Omega(\omega)\right|_{CI}}$ as a dashed curve in bottom panel of Fig.~\ref{fig:populations}.
When this ratio is below $\sim0.1$ the sequential model captures the important physics of the process, but as the ratio increases the sequential model no longer describes the simulation accurately.

With an intuitive understanding of the result of our simulation we turn our attention to the prediction made by this simulation.
Figure~\ref{fig:CompareEV} shows how the control field strength, $F$, and initial wavepacket position, $x_0$, affect the control dynamics.
The optimal photon energy for producing excited state population has only a slight dependence on the field intensity.
For higher field strength, the population of the excited state increases for all photon energies, and the optimal photon energy shifts slightly toward higher photon energy.
This is captured in the sequential model, eqn~\ref{TotalProb}; for higher field strength both $P_L$ and $P_R$ increase, because $\bar{\Omega}$ increases with field strength, leading to more population transfer.
The position of the optimal photon energy for excited state population also has a dependence on the initial wavepacket position.
According to the sequential model, the amount of population transferred is a competition between the velocity of the wavepacket at the CI and the velocity of the wavepacket in the resonance region.
This balance is changed by varying $x_0$ since it affects the wavepacket velocity at each of these points.

The technique demonstrated here is a way to control the final population of the decay pathways, and it also provides an avenue for controlling the shape of the resulting wavepacket.  
The total effect is that there is an  optimal photon energy for population transfer to the excited state.  
In the case of $\unitfrac[10^9]{V}{m}$, $\unit[0.5]{eV}$ is the most effective photon energy to maintain population on the excited adiabatic state.  

\subsection{Geometric phase control}

The previous section considered control of the molecular decay pathway in a continuous light-field.
The results of our simulation were easily explained using a semi-classical model which considers ``classical'' trajectories for the wavepacket. 
In this section we describe the use of a pulsed light-field applied after the wavepacket exits the conical intersection, as shown in Figure~\ref{fig:GPC}.
After traversing the CI, the wavepacket is split, and there will be population in both the excited and ground electronic states.
The control-field will couple these two parts of the wavepacket, and depending on the relative phase between them, this coupling will lead to constructive or destructive interference in the wavepacket. 
In contrast to the kinetic energy control mechanism, the ability to exert control in this situation results from the coherence prepared by the conical intersection, and this can only be described quantum mechanically.  
 
Consider the simple linear vibronic coupling diabatic potential given by eqn.~\ref{Potential}. If all of the wavefunction starts on the excited diabatic state, then any population transferred to the ground diabatic state will do so because of the linear coupling $\lambda y$.  
This coupling is negative for $y<0$ and positive for $y>0$.
The linearity of the coupling in the $y$-coordinate results in a $\pi$-phase jump at $y=0$ for the part of the wavepacket which stays on the adiabat after traversing the CI. 
The part of the wavepacket which stays on the diabat, i.e. undergoes a non-adiabatic transition, will not accumulate this phase.  
Comparing the two parts of the wavepacket created by traversing the CI, for $y > 0$ the upper and lower state wavepackets will have no phase difference, and for $y < 0$ the two wavepackets will have a $\pi$ phase difference.  
As a result, interfering the two wavepackets will generate an asymmetric wavepacket about $y = 0$.
This $y$-dependent phase-shift is directly related to the geometric phase, hence the name, ``geometric phase control."

\begin{figure}
	%\centering
	\includegraphics[width = \columnwidth]{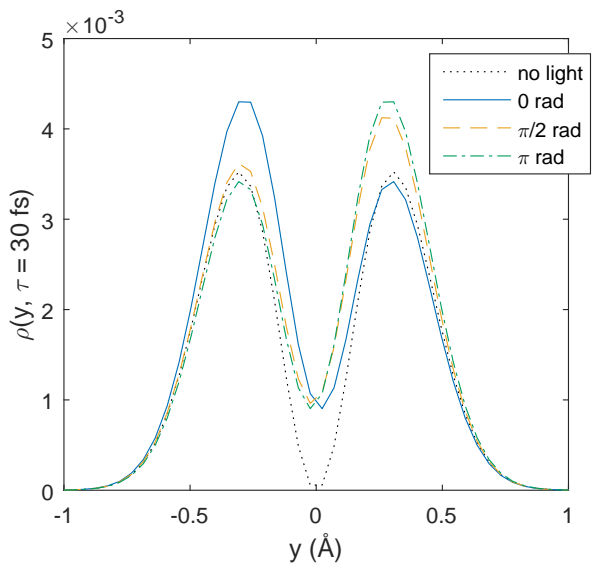}
    \caption{$y$-dependent probability density, $\rho(y,\tau=\unit[30]{fs})=\int\int dx dz \left|\psi(\tau=\unit[30]{fs})\right|^2$, of the excited adiabatic state. Depending on the phase of the applied light field, it is possible to constructively or destructively interfere the two parts of the nuclear wavepacket generated by a conical intersection to produce asymmetric distributions in the linear coupling direction.}
    \label{fig:Interfere}
\end{figure}

In Fig.~\ref{fig:Interfere} we show the $y$-dependent probability density, $\rho(y)=\int\int dx dz \left|\psi(\tau)\right|^2$, of the excited adiabatic state.
In addition to the field-free case~(black dotted line), we show the probability density following a sine-squared laser pulse that is on for $\unit[6]{fs}$ with a central photon energy of $\unit[1]{eV}$ and different values for the carrier envelope phase~(CEP). 
Changing the pulse-width of the control field has little effect on the probability density, so long as the two parts of the wavepacket on the different surfaces (which have a different kinetic energy) still have sufficient overlap.
To avoid overlap effects, we use a short pulse duration,  time-delayed by $\unit[24]{fs}$ from the start of the wavepacket propagation so that a majority of the wavepacket has passed through the conical intersection and the center of the wavepacket is in resonance.  
Again a peak field strength $F = \unitfrac[10^9]{V}{m}$ is used, with the same potentials as in the previous control scheme, however the linear coupling $\lambda$ is increased to by a factor of 10 so that more population is maintained on the excited adiabatic state.
Figure~\ref{fig:Interfere} shows that the asymmetry in the wavepacket along the linear coupling direction~$(y)$ depends on the phase of the field.
By changing the phase of the carrier from 0 to $\pi$ radians, the effect can be reversed.  
However, because the two wavepackets are moving relative to one another and changing their relative phase at each position, a phase of $\nicefrac{\pi}{2}$ does not entirely eliminate the asymmetry because the wavepacket only has significant interference during one half of the cycle.

\begin{figure}[t]
	\centering
    \includegraphics[width = .5\textwidth]{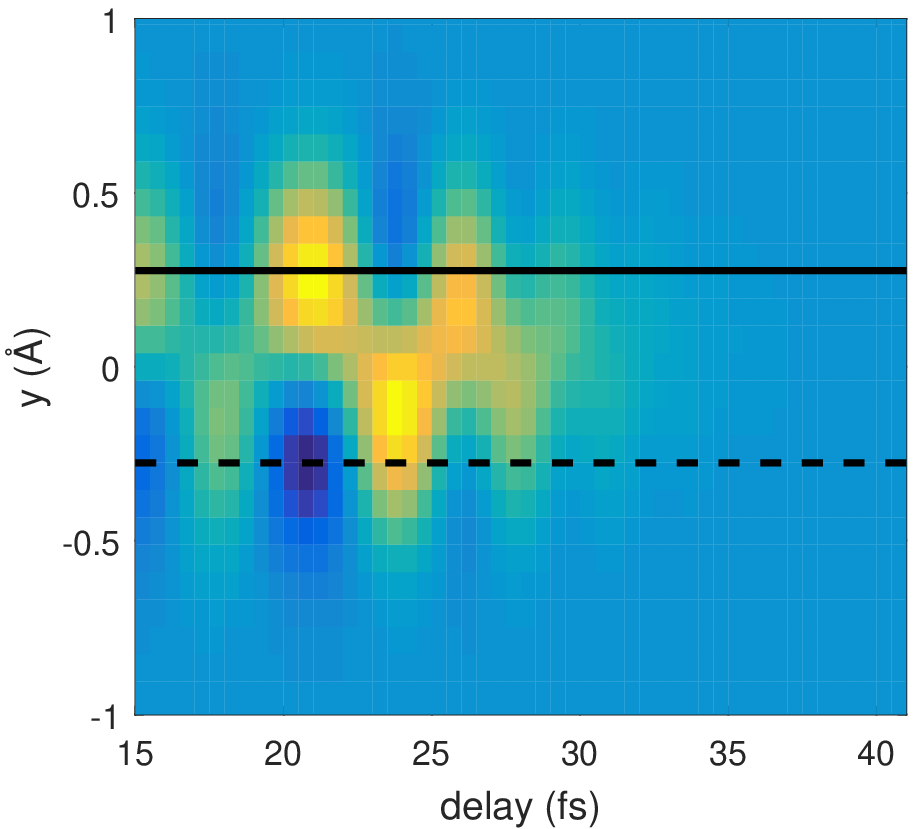}
    \includegraphics[width = 0.5\textwidth]{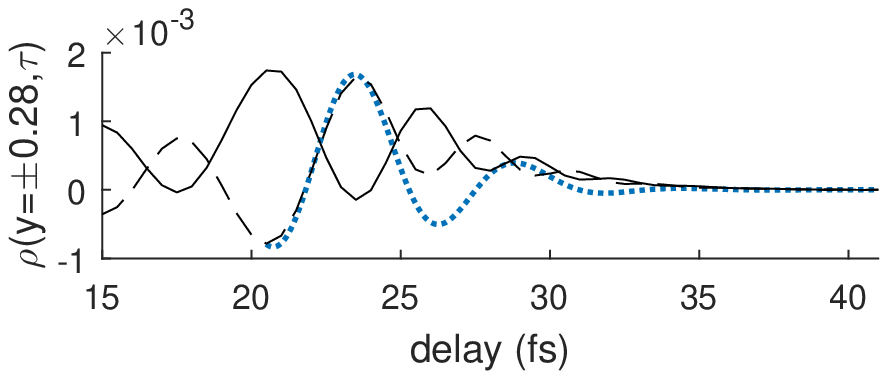}
    \caption{The top panel shows the delay dependence of $y$-dependent probability density, $\rho(y,\tau)=\int\int dx dz \left|\psi(\tau)\right|^2$ of the excited adiabatic state. The control pulse has a $\unit[6]{fs}$ duration and $\unit[1]{eV}$ central frequency. The result at the longest delay where the two wavepackets are well separated and do not interfere (\unit[41]{fs}) is subtracted to show the change more clearly. The bottom panel shows the delay-dependent probability density along the red and blue lines in the top panel~($y\sim\pm0.18$). The dotted blue curve is a damped sinusoidal function to guide the eye and accentuate the delay-dependent period of the oscillation. }    
    \label{fig:ChangeDelay}
\end{figure}

In order to map out this effect, we scan the time-delay of the control pulse in Fig.~\ref{fig:ChangeDelay}, which shows how the probability density changes with the delay of the control pulse. 
As the two wavepackets move relative to one another, their overall phase changes, and asymmetry of the wavepacket oscillates.  
Such an experiment is easily realized in strong-field driven high harmonic generation~(HHG), where the driving laser pulse can be separated from the XUV pulses and time-delayed, while maintaining definite relationship between the XUV envelope and the phase of the driving IR field.    
Delay-dependent line-outs of the probability density along $y=\pm 0.28$\AA~are shown in Fig.~\ref{fig:ChangeDelay}.  
Along these cuts, the probability density shows a decaying oscillatory behavior with a decreasing oscillation frequency.  
As the two parts of the wavepacket leave the vicinity of the CI, the energy separation between the states begins to increase and the phase velocities of the parts of the wavepacket begin to walk-off from one another. 
As the time-delay increases, the phase velocity difference increases and leads to faster oscillation between constructive and destructive interference as the phase-fronts move past one another. 

The decay of the interference signal represents a loss of electronic coherence due to a group velocity mismatch between the two parts of the wavepacket.
Currently, there is no nuclear wavepacket dispersion considered in this model, so, the electronic coherence will revive once the parts of the nuclear wavepacket overlap again. 
The loss of electronic coherence due to nuclear wavepacket walk off has been discussed in literature~\cite{halasz_nuclear-wave-packet_2013,mendive-tapia_coupled_2013,vacher_electron_2015} and our simple model suggests a possible protocol which could be used to study the complex interplay between electronic coherence and nuclear motion in the laboratory.
A wavepacket initially launched on an excited electronic states propagates toward a CI.  At the CI, the wavepacket bifurcates into two different electronic states.
The coherence between the electronic states can be monitored using a time-delayed control-field, as considered here. 
So in addition to a method for examining the geometric phase and exerting control of the wavepacket in the linear coupling direction, this protocol can also be used to investigate the effect of nuclear motion on electronic coherence. 

\section{Conclusion}
Conical intersections are the dominant mechanism for ultrafast, non-radiative relaxation in molecular systems.
This is due to the large non-adiabatic coupling that exists around CI.
In addition to the coupling provided by the CI, moderate intensity laser fields can also create couplings between the molecular states.
We have constructed a model to explore how these two different kinds of couplings can work together to control quantum evolution.  

A first example shows that population in different decay pathways can be controlled by choosing the photon energy of an applied continuous light field.  
The photon energy controls the probability of population transfer at resonant points between the two surfaces.  
Experiments can look for changes of decay pathway populations by tuning the photon energy.  
These experiments will also provide information about the shape of the potential energy surfaces and onset of non-adiabatic couplings.

A second example of control uses the phase difference accumulated by a wavepacket split by a conical intersection.  
We demonstrated that the diabatic linear coupling model predicts that a light field applied after the wavepacket emerges from the CI can be used to generate an asymmetric wavepacket on either side of the linear coupling vector.  
In this study we only consider a relatively simple light pulse with a single central frequency.  
More interesting light fields could be applied, such as a chirped pulse, to change the range over which coupling occurs, to allow for a range of resulting wavepackets.  
By generating wavepackets of different shapes, it might be possible (depending on the linear coupling direction) to look for changes in geometry which could be detected by coulomb explosion or similar techniques.
Moreover, this control protocol provides an interesting way to study to interplay between electronic coherence and nuclear motion.  

\section{Acknowledgments}

We would like to thank both Zheng Li and Todd Martinez of the PULSE Institute and the Stanford Chemistry Department and Oriol Vendrell at CFEL-DESY and the Hamburg Center for Ultrafast Imaging for their invaluable insight into this topic.  This work was supported by the National Science Foundation under Grant No. PHY-0649578.  CL-S was supported by the DOE SCGSR fellowship program.  JPC is supported by the U.S. Department of Energy (DOE), Office of Science, Basic Energy Sciences (BES), Chemical Sciences, Geosciences, and Biosciences Division.  

\bibliography{Zotero}

\begin{thebibliography}{26}%
\makeatletter
\providecommand \@ifxundefined [1]{%
 \@ifx{#1\undefined}
}%
\providecommand \@ifnum [1]{%
 \ifnum #1\expandafter \@firstoftwo
 \else \expandafter \@secondoftwo
 \fi
}%
\providecommand \@ifx [1]{%
 \ifx #1\expandafter \@firstoftwo
 \else \expandafter \@secondoftwo
 \fi
}%
\providecommand \natexlab [1]{#1}%
\providecommand \enquote  [1]{``#1''}%
\providecommand \bibnamefont  [1]{#1}%
\providecommand \bibfnamefont [1]{#1}%
\providecommand \citenamefont [1]{#1}%
\providecommand \href@noop [0]{\@secondoftwo}%
\providecommand \href [0]{\begingroup \@sanitize@url \@href}%
\providecommand \@href[1]{\@@startlink{#1}\@@href}%
\providecommand \@@href[1]{\endgroup#1\@@endlink}%
\providecommand \@sanitize@url [0]{\catcode `\\12\catcode `\$12\catcode
  `\&12\catcode `\#12\catcode `\^12\catcode `\_12\catcode `\%12\relax}%
\providecommand \@@startlink[1]{}%
\providecommand \@@endlink[0]{}%
\providecommand \url  [0]{\begingroup\@sanitize@url \@url }%
\providecommand \@url [1]{\endgroup\@href {#1}{\urlprefix }}%
\providecommand \urlprefix  [0]{URL }%
\providecommand \Eprint [0]{\href }%
\providecommand \doibase [0]{http://dx.doi.org/}%
\providecommand \selectlanguage [0]{\@gobble}%
\providecommand \bibinfo  [0]{\@secondoftwo}%
\providecommand \bibfield  [0]{\@secondoftwo}%
\providecommand \translation [1]{[#1]}%
\providecommand \BibitemOpen [0]{}%
\providecommand \bibitemStop [0]{}%
\providecommand \bibitemNoStop [0]{.\EOS\space}%
\providecommand \EOS [0]{\spacefactor3000\relax}%
\providecommand \BibitemShut  [1]{\csname bibitem#1\endcsname}%
\let\auto@bib@innerbib\@empty
%</preamble>
\bibitem [{\citenamefont {Dehareng}\ \emph {et~al.}(1983)\citenamefont
  {Dehareng}, \citenamefont {Chapuisat}, \citenamefont {Lorquet}, \citenamefont
  {Galloy},\ and\ \citenamefont {Raseev}}]{dehareng_dynamical_1983}%
  \BibitemOpen
  \bibfield  {author} {\bibinfo {author} {\bibfnamefont {D.}~\bibnamefont
  {Dehareng}}, \bibinfo {author} {\bibfnamefont {X.}~\bibnamefont {Chapuisat}},
  \bibinfo {author} {\bibfnamefont {J.-C.}\ \bibnamefont {Lorquet}}, \bibinfo
  {author} {\bibfnamefont {C.}~\bibnamefont {Galloy}}, \ and\ \bibinfo {author}
  {\bibfnamefont {G.}~\bibnamefont {Raseev}},\ }\href {\doibase
  10.1063/1.444862} {\bibfield  {journal} {\bibinfo  {journal} {The Journal of
  Chemical Physics}\ }\textbf {\bibinfo {volume} {78}},\ \bibinfo {pages}
  {1246} (\bibinfo {year} {1983})}\BibitemShut {NoStop}%
\bibitem [{\citenamefont {Levine}\ and\ \citenamefont
  {Martínez}(2007)}]{levine_isomerization_2007}%
  \BibitemOpen
  \bibfield  {author} {\bibinfo {author} {\bibfnamefont {B.~G.}\ \bibnamefont
  {Levine}}\ and\ \bibinfo {author} {\bibfnamefont {T.~J.}\ \bibnamefont
  {Martínez}},\ }\href {\doibase 10.1146/annurev.physchem.57.032905.104612}
  {\bibfield  {journal} {\bibinfo  {journal} {Annual Review of Physical
  Chemistry}\ }\textbf {\bibinfo {volume} {58}},\ \bibinfo {pages} {613}
  (\bibinfo {year} {2007})}\BibitemShut {NoStop}%
\bibitem [{\citenamefont {Musser}\ \emph {et~al.}(2015)\citenamefont {Musser},
  \citenamefont {Liebel}, \citenamefont {Schnedermann}, \citenamefont {Wende},
  \citenamefont {Kehoe}, \citenamefont {Rao},\ and\ \citenamefont
  {Kukura}}]{musser_evidence_2015}%
  \BibitemOpen
  \bibfield  {author} {\bibinfo {author} {\bibfnamefont {A.~J.}\ \bibnamefont
  {Musser}}, \bibinfo {author} {\bibfnamefont {M.}~\bibnamefont {Liebel}},
  \bibinfo {author} {\bibfnamefont {C.}~\bibnamefont {Schnedermann}}, \bibinfo
  {author} {\bibfnamefont {T.}~\bibnamefont {Wende}}, \bibinfo {author}
  {\bibfnamefont {T.~B.}\ \bibnamefont {Kehoe}}, \bibinfo {author}
  {\bibfnamefont {A.}~\bibnamefont {Rao}}, \ and\ \bibinfo {author}
  {\bibfnamefont {P.}~\bibnamefont {Kukura}},\ }\href {\doibase
  10.1038/nphys3241} {\bibfield  {journal} {\bibinfo  {journal} {Nature
  Physics}\ }\textbf {\bibinfo {volume} {11}},\ \bibinfo {pages} {352}
  (\bibinfo {year} {2015})}\BibitemShut {NoStop}%
\bibitem [{\citenamefont {Klessinger}\ and\ \citenamefont
  {Michl}(1995)}]{klessinger_excited_1995}%
  \BibitemOpen
  \bibfield  {author} {\bibinfo {author} {\bibfnamefont {M.}~\bibnamefont
  {Klessinger}}\ and\ \bibinfo {author} {\bibfnamefont {J.}~\bibnamefont
  {Michl}},\ }\href@noop {} {\emph {\bibinfo {title} {Excited states and
  photochemistry of organic molecules}}}\ (\bibinfo  {publisher} {VCH},\
  \bibinfo {address} {New York},\ \bibinfo {year} {1995})\ \bibinfo {note}
  {oCLC: 27226211}\BibitemShut {NoStop}%
\bibitem [{\citenamefont {Allen}\ and\ \citenamefont
  {{ICCS}}(2009)}]{allen_computational_2009}%
  \BibitemOpen
  \bibinfo {editor} {\bibfnamefont {G.}~\bibnamefont {Allen}}\ and\ \bibinfo
  {editor} {\bibnamefont {{ICCS}}},\ eds.,\ \href@noop {} {\emph {\bibinfo
  {title} {Computational science - {ICCS} 2009:: 9th international conference,
  {Baton} {Rouge}, {LA}, {USA}, {May} 25-27, 2009; proceedings. {Pt}. 2:
  [...]}}},\ \bibinfo {series} {Lecture notes in computer science}\ No.\
  \bibinfo {number} {5545}\ (\bibinfo  {publisher} {Springer},\ \bibinfo
  {address} {Berlin},\ \bibinfo {year} {2009})\ \bibinfo {note} {oCLC:
  552137512}\BibitemShut {NoStop}%
\bibitem [{\citenamefont {Born}\ and\ \citenamefont
  {Oppenheimer}(1927)}]{born_zur_1927}%
  \BibitemOpen
  \bibfield  {author} {\bibinfo {author} {\bibfnamefont {M.}~\bibnamefont
  {Born}}\ and\ \bibinfo {author} {\bibfnamefont {R.}~\bibnamefont
  {Oppenheimer}},\ }\href {\doibase 10.1002/andp.19273892002} {\bibfield
  {journal} {\bibinfo  {journal} {Annalen der Physik}\ }\textbf {\bibinfo
  {volume} {389}},\ \bibinfo {pages} {457} (\bibinfo {year}
  {1927})}\BibitemShut {NoStop}%
\bibitem [{\citenamefont {von Neumann}\ and\ \citenamefont
  {Wigner}(1929)}]{von_neumann_uber_1929}%
  \BibitemOpen
  \bibfield  {author} {\bibinfo {author} {\bibfnamefont {J.}~\bibnamefont {von
  Neumann}}\ and\ \bibinfo {author} {\bibfnamefont {E.}~\bibnamefont
  {Wigner}},\ }\href@noop {} {\bibfield  {journal} {\bibinfo  {journal}
  {Physikalische Zeitschrift}\ }\textbf {\bibinfo {volume} {30}},\ \bibinfo
  {pages} {467} (\bibinfo {year} {1929})}\BibitemShut {NoStop}%
\bibitem [{\citenamefont {Baer}(2006)}]{baer_beyond_2006}%
  \BibitemOpen
  \bibfield  {author} {\bibinfo {author} {\bibfnamefont {M.}~\bibnamefont
  {Baer}},\ }\href@noop {} {\emph {\bibinfo {title} {Beyond
  {Born}-{Oppenheimer}: conical intersections and electronic non-adiabatic
  coupling terms}}}\ (\bibinfo  {publisher} {Wiley},\ \bibinfo {address}
  {Hoboken, N.J.},\ \bibinfo {year} {2006})\BibitemShut {NoStop}%
\bibitem [{\citenamefont {Worth}\ and\ \citenamefont
  {Cederbaum}(2004)}]{worth_beyond_2004}%
  \BibitemOpen
  \bibfield  {author} {\bibinfo {author} {\bibfnamefont {G.~A.}\ \bibnamefont
  {Worth}}\ and\ \bibinfo {author} {\bibfnamefont {L.~S.}\ \bibnamefont
  {Cederbaum}},\ }\href {\doibase 10.1146/annurev.physchem.55.091602.094335}
  {\bibfield  {journal} {\bibinfo  {journal} {Annual Review of Physical
  Chemistry}\ }\textbf {\bibinfo {volume} {55}},\ \bibinfo {pages} {127}
  (\bibinfo {year} {2004})}\BibitemShut {NoStop}%
\bibitem [{\citenamefont {Yarkony}(1996)}]{yarkony_diabolical_1996}%
  \BibitemOpen
  \bibfield  {author} {\bibinfo {author} {\bibfnamefont {D.~R.}\ \bibnamefont
  {Yarkony}},\ }\href {\doibase 10.1103/RevModPhys.68.985} {\bibfield
  {journal} {\bibinfo  {journal} {Reviews of Modern Physics}\ }\textbf
  {\bibinfo {volume} {68}},\ \bibinfo {pages} {985} (\bibinfo {year}
  {1996})}\BibitemShut {NoStop}%
\bibitem [{\citenamefont {Domcke}\ \emph {et~al.}(2004)\citenamefont {Domcke},
  \citenamefont {Yarkony},\ and\ \citenamefont
  {Köppel}}]{domcke_conical_2004}%
  \BibitemOpen
  \bibinfo {editor} {\bibfnamefont {W.}~\bibnamefont {Domcke}}, \bibinfo
  {editor} {\bibfnamefont {D.}~\bibnamefont {Yarkony}}, \ and\ \bibinfo
  {editor} {\bibfnamefont {H.}~\bibnamefont {Köppel}},\ eds.,\ \href@noop {}
  {\emph {\bibinfo {title} {Conical intersections: electronic structure,
  dynamics \& spectroscopy}}},\ \bibinfo {series} {Advanced series in physical
  chemistry}\ No.\ \bibinfo {number} {v.15}\ (\bibinfo  {publisher} {World
  Scientific},\ \bibinfo {address} {River Edge, NJ},\ \bibinfo {year} {2004})\
  \bibinfo {note} {oCLC: ocm56885304}\BibitemShut {NoStop}%
\bibitem [{\citenamefont {Van~Voorhis}\ \emph {et~al.}(2010)\citenamefont
  {Van~Voorhis}, \citenamefont {Kowalczyk}, \citenamefont {Kaduk},
  \citenamefont {Wang}, \citenamefont {Cheng},\ and\ \citenamefont
  {Wu}}]{van_voorhis_diabatic_2010}%
  \BibitemOpen
  \bibfield  {author} {\bibinfo {author} {\bibfnamefont {T.}~\bibnamefont
  {Van~Voorhis}}, \bibinfo {author} {\bibfnamefont {T.}~\bibnamefont
  {Kowalczyk}}, \bibinfo {author} {\bibfnamefont {B.}~\bibnamefont {Kaduk}},
  \bibinfo {author} {\bibfnamefont {L.-P.}\ \bibnamefont {Wang}}, \bibinfo
  {author} {\bibfnamefont {C.-L.}\ \bibnamefont {Cheng}}, \ and\ \bibinfo
  {author} {\bibfnamefont {Q.}~\bibnamefont {Wu}},\ }\href {\doibase
  10.1146/annurev.physchem.012809.103324} {\bibfield  {journal} {\bibinfo
  {journal} {Annual Review of Physical Chemistry}\ }\textbf {\bibinfo {volume}
  {61}},\ \bibinfo {pages} {149} (\bibinfo {year} {2010})}\BibitemShut
  {NoStop}%
\bibitem [{\citenamefont {Teller}(1937)}]{teller_crossing_1937}%
  \BibitemOpen
  \bibfield  {author} {\bibinfo {author} {\bibfnamefont {E.}~\bibnamefont
  {Teller}},\ }\href {\doibase 10.1021/j150379a010} {\bibfield  {journal}
  {\bibinfo  {journal} {The Journal of Physical Chemistry}\ }\textbf {\bibinfo
  {volume} {41}},\ \bibinfo {pages} {109} (\bibinfo {year} {1937})}\BibitemShut
  {NoStop}%
\bibitem [{\citenamefont {Domcke}\ \emph {et~al.}(2011)\citenamefont {Domcke},
  \citenamefont {Yarkony},\ and\ \citenamefont
  {Köppel}}]{domcke_conical_2011}%
  \BibitemOpen
  \bibinfo {editor} {\bibfnamefont {W.}~\bibnamefont {Domcke}}, \bibinfo
  {editor} {\bibfnamefont {D.}~\bibnamefont {Yarkony}}, \ and\ \bibinfo
  {editor} {\bibfnamefont {H.}~\bibnamefont {Köppel}},\ eds.,\ \href@noop {}
  {\emph {\bibinfo {title} {Conical intersections: theory, computation and
  experiment}}},\ \bibinfo {series} {Advanced series in physical chemistry}\
  No.\ \bibinfo {number} {v. 17}\ (\bibinfo  {publisher} {World Scientific},\
  \bibinfo {address} {Singapore ; Hackensack, NJ},\ \bibinfo {year} {2011})\
  \bibinfo {note} {oCLC: ocn607976450}\BibitemShut {NoStop}%
\bibitem [{\citenamefont {Kowalewski}\ \emph {et~al.}(2015)\citenamefont
  {Kowalewski}, \citenamefont {Bennett}, \citenamefont {Dorfman},\ and\
  \citenamefont {Mukamel}}]{kowalewski_catching_2015}%
  \BibitemOpen
  \bibfield  {author} {\bibinfo {author} {\bibfnamefont {M.}~\bibnamefont
  {Kowalewski}}, \bibinfo {author} {\bibfnamefont {K.}~\bibnamefont {Bennett}},
  \bibinfo {author} {\bibfnamefont {K.~E.}\ \bibnamefont {Dorfman}}, \ and\
  \bibinfo {author} {\bibfnamefont {S.}~\bibnamefont {Mukamel}},\ }\href
  {\doibase 10.1103/PhysRevLett.115.193003} {\bibfield  {journal} {\bibinfo
  {journal} {Physical Review Letters}\ }\textbf {\bibinfo {volume} {115}}
  (\bibinfo {year} {2015}),\ 10.1103/PhysRevLett.115.193003}\BibitemShut
  {NoStop}%
\bibitem [{\citenamefont {Kim}\ \emph {et~al.}(2012)\citenamefont {Kim},
  \citenamefont {Tao}, \citenamefont {White}, \citenamefont {Petrović},
  \citenamefont {Martinez},\ and\ \citenamefont
  {Bucksbaum}}]{kim_control_2012}%
  \BibitemOpen
  \bibfield  {author} {\bibinfo {author} {\bibfnamefont {J.}~\bibnamefont
  {Kim}}, \bibinfo {author} {\bibfnamefont {H.}~\bibnamefont {Tao}}, \bibinfo
  {author} {\bibfnamefont {J.~L.}\ \bibnamefont {White}}, \bibinfo {author}
  {\bibfnamefont {V.~S.}\ \bibnamefont {Petrović}}, \bibinfo {author}
  {\bibfnamefont {T.~J.}\ \bibnamefont {Martinez}}, \ and\ \bibinfo {author}
  {\bibfnamefont {P.~H.}\ \bibnamefont {Bucksbaum}},\ }\href {\doibase
  10.1021/jp208384b} {\bibfield  {journal} {\bibinfo  {journal} {The Journal of
  Physical Chemistry A}\ }\textbf {\bibinfo {volume} {116}},\ \bibinfo {pages}
  {2758} (\bibinfo {year} {2012})}\BibitemShut {NoStop}%
\bibitem [{\citenamefont {Sussman}\ \emph {et~al.}(2006)\citenamefont
  {Sussman}, \citenamefont {Townsend}, \citenamefont {Ivanov},\ and\
  \citenamefont {Stolow}}]{sussman_dynamic_2006}%
  \BibitemOpen
  \bibfield  {author} {\bibinfo {author} {\bibfnamefont {B.~J.}\ \bibnamefont
  {Sussman}}, \bibinfo {author} {\bibfnamefont {D.}~\bibnamefont {Townsend}},
  \bibinfo {author} {\bibfnamefont {M.~Y.}\ \bibnamefont {Ivanov}}, \ and\
  \bibinfo {author} {\bibfnamefont {A.}~\bibnamefont {Stolow}},\ }\href
  {\doibase 10.1126/science.1132289} {\bibfield  {journal} {\bibinfo  {journal}
  {Science}\ }\textbf {\bibinfo {volume} {314}},\ \bibinfo {pages} {278}
  (\bibinfo {year} {2006})}\BibitemShut {NoStop}%
\bibitem [{\citenamefont {Townsend}\ \emph {et~al.}(2011)\citenamefont
  {Townsend}, \citenamefont {Sussman},\ and\ \citenamefont
  {Stolow}}]{townsend_stark_2011}%
  \BibitemOpen
  \bibfield  {author} {\bibinfo {author} {\bibfnamefont {D.}~\bibnamefont
  {Townsend}}, \bibinfo {author} {\bibfnamefont {B.~J.}\ \bibnamefont
  {Sussman}}, \ and\ \bibinfo {author} {\bibfnamefont {A.}~\bibnamefont
  {Stolow}},\ }\href {\doibase 10.1021/jp109095d} {\bibfield  {journal}
  {\bibinfo  {journal} {The Journal of Physical Chemistry A}\ }\textbf
  {\bibinfo {volume} {115}},\ \bibinfo {pages} {357} (\bibinfo {year}
  {2011})}\BibitemShut {NoStop}%
\bibitem [{\citenamefont {Sussman}\ \emph {et~al.}(2005)\citenamefont
  {Sussman}, \citenamefont {Ivanov},\ and\ \citenamefont
  {Stolow}}]{sussman_nonperturbative_2005}%
  \BibitemOpen
  \bibfield  {author} {\bibinfo {author} {\bibfnamefont {B.~J.}\ \bibnamefont
  {Sussman}}, \bibinfo {author} {\bibfnamefont {M.~Y.}\ \bibnamefont {Ivanov}},
  \ and\ \bibinfo {author} {\bibfnamefont {A.}~\bibnamefont {Stolow}},\ }\href
  {\doibase 10.1103/PhysRevA.71.051401} {\bibfield  {journal} {\bibinfo
  {journal} {Physical Review A}\ }\textbf {\bibinfo {volume} {71}} (\bibinfo
  {year} {2005}),\ 10.1103/PhysRevA.71.051401}\BibitemShut {NoStop}%
\bibitem [{\citenamefont {Althorpe}\ \emph {et~al.}(2008)\citenamefont
  {Althorpe}, \citenamefont {Stecher},\ and\ \citenamefont
  {Bouakline}}]{althorpe_effect_2008}%
  \BibitemOpen
  \bibfield  {author} {\bibinfo {author} {\bibfnamefont {S.~C.}\ \bibnamefont
  {Althorpe}}, \bibinfo {author} {\bibfnamefont {T.}~\bibnamefont {Stecher}}, \
  and\ \bibinfo {author} {\bibfnamefont {F.}~\bibnamefont {Bouakline}},\ }\href
  {\doibase 10.1063/1.3031215} {\bibfield  {journal} {\bibinfo  {journal} {The
  Journal of Chemical Physics}\ }\textbf {\bibinfo {volume} {129}},\ \bibinfo
  {pages} {214117} (\bibinfo {year} {2008})}\BibitemShut {NoStop}%
\bibitem [{\citenamefont {Tannor}(2007)}]{tannor_introduction_2007}%
  \BibitemOpen
  \bibfield  {author} {\bibinfo {author} {\bibfnamefont {D.~J.}\ \bibnamefont
  {Tannor}},\ }\href@noop {} {\emph {\bibinfo {title} {Introduction to quantum
  mechanics: a time-dependent perspective}}}\ (\bibinfo  {publisher}
  {University Science Books},\ \bibinfo {address} {Sausalito, Calif},\ \bibinfo
  {year} {2007})\ \bibinfo {note} {oCLC: ocm65207147}\BibitemShut {NoStop}%
\bibitem [{\citenamefont {Kosloff}\ and\ \citenamefont
  {Kosloff}(1983)}]{kosloff_fourier_1983}%
  \BibitemOpen
  \bibfield  {author} {\bibinfo {author} {\bibfnamefont {D.}~\bibnamefont
  {Kosloff}}\ and\ \bibinfo {author} {\bibfnamefont {R.}~\bibnamefont
  {Kosloff}},\ }\href {\doibase 10.1016/0021-9991(83)90015-3} {\bibfield
  {journal} {\bibinfo  {journal} {Journal of Computational Physics}\ }\textbf
  {\bibinfo {volume} {52}},\ \bibinfo {pages} {35} (\bibinfo {year}
  {1983})}\BibitemShut {NoStop}%
\bibitem [{\citenamefont {Landau}\ \emph {et~al.}(1958)\citenamefont {Landau},
  \citenamefont {Lifshitz}, \citenamefont {Sykes}, \citenamefont {Bell},\ and\
  \citenamefont {Rose}}]{landau_quantum_1958}%
  \BibitemOpen
  \bibfield  {author} {\bibinfo {author} {\bibfnamefont {L.~D.}\ \bibnamefont
  {Landau}}, \bibinfo {author} {\bibfnamefont {E.~M.}\ \bibnamefont
  {Lifshitz}}, \bibinfo {author} {\bibfnamefont {J.~B.}\ \bibnamefont {Sykes}},
  \bibinfo {author} {\bibfnamefont {J.~S.}\ \bibnamefont {Bell}}, \ and\
  \bibinfo {author} {\bibfnamefont {M.~E.}\ \bibnamefont {Rose}},\ }\href
  {\doibase 10.1063/1.3062347} {\bibfield  {journal} {\bibinfo  {journal}
  {Physics Today}\ }\textbf {\bibinfo {volume} {11}},\ \bibinfo {pages} {56}
  (\bibinfo {year} {1958})}\BibitemShut {NoStop}%
\bibitem [{\citenamefont {Halász}\ \emph {et~al.}(2013)\citenamefont
  {Halász}, \citenamefont {Vibók}, \citenamefont {Moiseyev},\ and\
  \citenamefont {Cederbaum}}]{halasz_nuclear-wave-packet_2013}%
  \BibitemOpen
  \bibfield  {author} {\bibinfo {author} {\bibfnamefont {G.~J.}\ \bibnamefont
  {Halász}}, \bibinfo {author} {\bibfnamefont {Ã.}~\bibnamefont {Vibók}},
  \bibinfo {author} {\bibfnamefont {N.}~\bibnamefont {Moiseyev}}, \ and\
  \bibinfo {author} {\bibfnamefont {L.~S.}\ \bibnamefont {Cederbaum}},\ }\href
  {\doibase 10.1103/PhysRevA.88.043413} {\bibfield  {journal} {\bibinfo
  {journal} {Physical Review A}\ }\textbf {\bibinfo {volume} {88}} (\bibinfo
  {year} {2013}),\ 10.1103/PhysRevA.88.043413}\BibitemShut {NoStop}%
\bibitem [{\citenamefont {Mendive-Tapia}\ \emph {et~al.}(2013)\citenamefont
  {Mendive-Tapia}, \citenamefont {Vacher}, \citenamefont {Bearpark},\ and\
  \citenamefont {Robb}}]{mendive-tapia_coupled_2013}%
  \BibitemOpen
  \bibfield  {author} {\bibinfo {author} {\bibfnamefont {D.}~\bibnamefont
  {Mendive-Tapia}}, \bibinfo {author} {\bibfnamefont {M.}~\bibnamefont
  {Vacher}}, \bibinfo {author} {\bibfnamefont {M.~J.}\ \bibnamefont
  {Bearpark}}, \ and\ \bibinfo {author} {\bibfnamefont {M.~A.}\ \bibnamefont
  {Robb}},\ }\href {\doibase 10.1063/1.4815914} {\bibfield  {journal} {\bibinfo
   {journal} {The Journal of Chemical Physics}\ }\textbf {\bibinfo {volume}
  {139}},\ \bibinfo {pages} {044110} (\bibinfo {year} {2013})}\BibitemShut
  {NoStop}%
\bibitem [{\citenamefont {Vacher}\ \emph {et~al.}(2015)\citenamefont {Vacher},
  \citenamefont {Steinberg}, \citenamefont {Jenkins}, \citenamefont
  {Bearpark},\ and\ \citenamefont {Robb}}]{vacher_electron_2015}%
  \BibitemOpen
  \bibfield  {author} {\bibinfo {author} {\bibfnamefont {M.}~\bibnamefont
  {Vacher}}, \bibinfo {author} {\bibfnamefont {L.}~\bibnamefont {Steinberg}},
  \bibinfo {author} {\bibfnamefont {A.~J.}\ \bibnamefont {Jenkins}}, \bibinfo
  {author} {\bibfnamefont {M.~J.}\ \bibnamefont {Bearpark}}, \ and\ \bibinfo
  {author} {\bibfnamefont {M.~A.}\ \bibnamefont {Robb}},\ }\href {\doibase
  10.1103/PhysRevA.92.040502} {\bibfield  {journal} {\bibinfo  {journal}
  {Physical Review A}\ }\textbf {\bibinfo {volume} {92}} (\bibinfo {year}
  {2015}),\ 10.1103/PhysRevA.92.040502}\BibitemShut {NoStop}%
\end{thebibliography}%

\end{document}